\documentclass[conference]{IEEEtranAAC2018}
\IEEEoverridecommandlockouts
\usepackage{hyperref}
\usepackage[noadjust]{cite}
\usepackage{amsmath,amssymb,amsfonts}
\usepackage{algorithmic}
\usepackage{graphicx}
\usepackage{textcomp}
\usepackage{xcolor}
\usepackage{balance}

\def\BibTeX{{\rm B\kern-.05em{\sc i\kern-.025em b}\kern-.08em
    T\kern-.1667em\lower.7ex\hbox{E}\kern-.125emX}}
		
\begin{document}
\bstctlcite{IEEEexample:BSTcontrol}
\title{All Optical Control of Beam Dynamics in a DLA\\
\thanks{This work was supported by Gordon and Betty Moore Foundation (GBMF4744), National Science Foundation (NSF) (PHY-1734215, PHY-1535711), and U.S. Department of Energy (DE-AC02-76SF00515, DE-SC0009914).}\thanks{Preprint prepared for the IEEE proceedings of Advanced Accelerator Concepts (AAC) 2018.}
}

\author{\IEEEauthorblockN{David Cesar}
\IEEEauthorblockA{\textit{Department of Physics and Astronomy}\\ UCLA, Los Angeles, CA, USA \\
Email: dcesar@ucla.edu}
\and
\IEEEauthorblockN{Pietro Musumeci}
\IEEEauthorblockA{\textit{Department of Physics and Astronomy}\\ UCLA, Los Angeles, CA, USA \\}
\and
\IEEEauthorblockN{Joel England}
\IEEEauthorblockA{\textit{SLAC National Accelerator Laboratory} \\ Menlo Park, CA, USA}}


%Telephone: (800) 555--1212, Fax: (888) 555--1212}


\maketitle

\begin{abstract}
Dielectric laser acceleration draws upon nano-fabrication techniques to build photonic structures for high gradient electron acceleration. At the small spatial scales characteristic of these structures conventional accelerator techniques become ineffective at stabilizing the beam dynamics. Instead we propose a scheme to stabilize the motion by directly modulating the drive laser, in analogy to a radio-frequency-quadrupole. Here we present a design for a programmable `lattice' being built at UCLA's Pegasus laboratory. The accelerator accepts an unmodulated 3.5 MeV electron beam and then bunches and  accelerates the beam by 1.5 MeV over a distance of 2 cm.
\end{abstract}

%  keywords

% \fi

\section{Introduction}
% State the topic of the paper
	Over the last three years the Accelerator on a Chip International Program (ACHIP)\,\cite{wootton_towards_2017} has made significant progress towards developing an integrated MeV scale accelerator based on dielectric laser acceleration (DLA)\,\cite{england_dielectric_2014}. For example, an accelerating gradient of 0.85 GeV/m and energy gain of 0.3 MeV have been demonstrated\,\cite{cesar_enhanced_2018,cesar_high-field_2018}, a realistic design for a multi-stage structure fed by a photonic waveguide network has been developed \cite{hughes_-chip_2018}, and  laser driven components for beam steering, staging, and focusing have been tested with sub-relativistic particles \cite{mcneur_elements_2018}. The next step is to demonstrate that an electron beam can be stably accelerated through the sub-micron vacuum channels intrinsic to DLA.
    
% Discuss why beam dynamics in DLA are unique
	From an electron's perspective the transverse magnetic (TM) mode in a DLA is similar to the TM mode in a conventional linac, provided the frequency is scaled by 50,000 times and the gradient by about 50 times. That scaling has major consequences. For example, both synchrotron and betatron frequencies increase proportionally to $\sqrt{E_0 \omega}$, where $E_0$ is the incident laser peak electric field and $\omega$ is the laser frequency. Consequently, at low beam energy the resonant electromagnetic defocusing force would overpower a 10\,MT/m quadrupole. Notably, however, the energy gain per period varies as $E_0/\omega$ so that at optical scales, the beam appears very stiff. This makes the beam dynamics in a DLA especially sensitive to changes in the phase of the accelerating mode, which we can use to control and confine the beam.     
    %(given by $\alpha = e G /\left(k_\text{res} m c^2\right)$)

% Compare solutions: RFQ, Naranjo, and APF [fix this paragraph]
	Since the defocusing force in a DLA is too large to overcome with magnetostatic focusing, we turn to all-electromagnetic methods to stabilize beam dynamics in field of a linac. The most well-known method is the radio-frequency quadrupole (RFQ), which physically modulates an RF cavity to create a non-synchronous quadrupole moment that then works like a FODO (focus-drift-focus-drift) lattice\,\cite{wangler_rf_2008}. Because our DLA devices have a planar symmetry, rather than cylindrical symmetry, it is impossible to exactly mimic the RFQ. However, we can force the beam to oscillate between longitudinally stable and transversely stable phases in analogy to the FODO lattice used in conventional accelerators. The `alternating phase focusing' scheme does exactly that\,\cite{niedermayer_alternating_2018}. 
    
\begin{figure}
\includegraphics[width=\columnwidth]{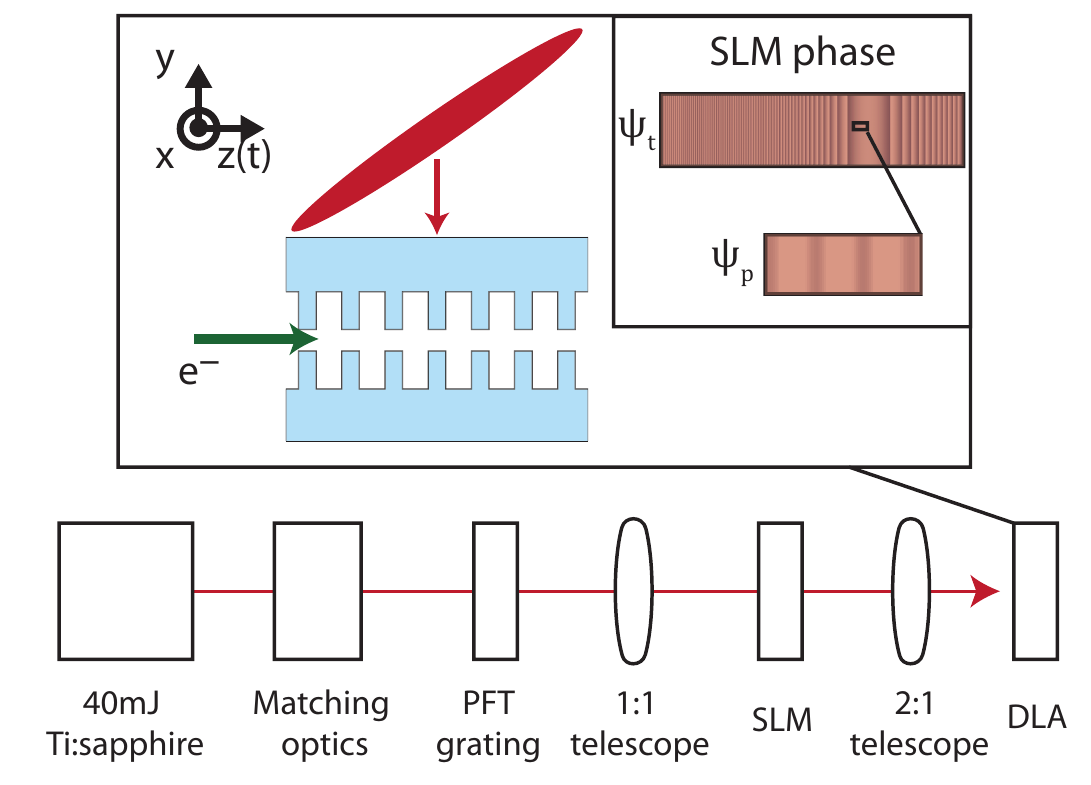}%
\caption{Cartoon of the proposed DLA experiment (not to scale) using a tailored phase profile imprinted onto a pulse front tilted laser beam to control the beam dynamics over a 2\,cm long structure. At bottom is a block diagram of the optical design. At top is a side view of the DLA interaction, including the $z$ dependent phase profile employed.}%
\label{fig:cartoon}
\end{figure}
    
    A more general solution to the problem of stability in a DLA was presented by Naranjo et al. as part of the GALAXIE project \cite{naranjo_stable_2012}. In their approach they simultaneously excite multiple modes of a DLA structure. One of the modes is synchronous and provides a linear accelerating force, while the others rapidly slip by the beam and create pondermotive focusing. They determine approximate stability criteria and show that by engineering the velocity and amplitude of each mode they can achieve stable acceleration for all beam energies. Then they applied this technique to an ambitious accelerator design\,\cite{naranjo_galaxie:_2013}, which is the inspiration for our present design. 
    
    %What all of these schemes have in common is that stable acceleration is created by adding a $z$ dependent phase to the accelerating mode.
    
    %Moreover, the parameters of the RFQ can be adiabatically adjusted so that accepts an unbunched beam and ejects an accerated, bunched

% Introduce phase mask & explain why it should work (Fig 1)
    We propose an accelerator design based on a programmable spatial light modulator (SLM) which is used to create multiple harmonics in a side-coupled DLA structure. Imaging an SLM onto the DLA allows us to program a $z$-dependent phase into the drive laser which will influence the beam dynamics as it illuminates the structure, as illustrated in Fig.\,(\ref{fig:cartoon}). We have already demonstrated a simpler form of optical phase control in a DLA by compensating for dephasing due to the Kerr effect\,\cite{cesar_high-field_2018}, and we have also demonstrated control of the phase velocity in a 1\,mm tilted pulse front DLA\,\cite{cesar_enhanced_2018}. Now we propose to elongate our optics to fit a 2\,cm structure and add the SLM so that we can stabilize the beam dynamics in real time.
%Unlike conventional linacs, which are resonant structures, DLA structures are broadband photonic devices\,\textemdash\,they must be in order to keep the pulse length short and avoid damaging the structure\,\textemdash\,which inherit the amplitude and phase of the drive laser pulse.     

%Summarize the content of this paper
	In these proceedings we discuss the beam dynamics of a DLA designed to bunch and accelerate part of the beam from UCLA's Pegasus facility\,\cite{maxson_direct_2017} over a distance of 2 cm. Based on the layout of Fig.\,(\ref{fig:cartoon}) we design a phase mask which uses the first centimeter to bunch the beam and the second to accelerate at an average gradient of 150\,MV/m for a total energy gain of 1.5\,MeV. We explain the design of the accelerator by applying approximate stability criteria derived from a simplified analysis of the beam dynamics.

%%%%%%%%%%%%%%%%%%%%%%%%%%%%%%%%%%%%%%%%%%%%%%%%%%%%%%%%%%%%%%%%%%%%%%%%%%%%%%%%%%
%%%%%%%%%%%%%%%%%%%%%%%%%%%%%%%%%%%%%%%%%%%%%%%%%%%%%%%%%%%%%%%%%%%%%%%%%%%%%%%%%%
%%%%%%%%%%%%%%%%%%%%%%%%%%%%%%%%%%%%%%%%%%%%%%%%%%%%%%%%%%%%%%%%%%%%%%%%%%%%%%%%%%
\section{Theory}
\label{sec:theory}

%Refer to naranjo&rosenzweig and discuss the meaning of resonant/non-res harmonics
Our study of beam dynamics is based on the formalism of Naranjo et al.\,\cite{naranjo_stable_2012} which uses the separation of scales technique to calculate the pondermotive focusing caused by a rapidly oscillating force. Explicitly, we decompose the electro-magnetic field into modes indexed by their phase velocity $\beta_m=\omega/c k_m$. Only one mode has the same velocity as the electron beam, and we use it to both accelerate and provide longitudinal stability. The non-resonant modes are used to provide transverse stability via pondermotive focusing.

% Show how harmonics are created (Jacobi-Anger)
Naranjo et.al. suggested that the non-resonant harmonics can be created by modulating the dielectric structure, however we find it more practical to generate them via a programmable phase mask. As illustrated in Fig.\,(\ref{fig:cartoon}) we can apply a sinusoidal phase modulation $a \cos(\psi_p)$ with $\psi_p \approx (\delta_k z)$ which is added to the Bloch phase factor of our periodic structure, $i k_g z$, so that the fields inside the structure will be proportional to:
\vspace{-.25\topskip}
\begin{equation}
	\label{eq:Jacobi_Anger}
    e^{i\left(k_g z +a \cos(\delta_k z )\right)}=\sum_{m=-\infty}^{\infty}i^m J_m(a)e^{i\left( (k_g+m\delta_k)z \right)}
\end{equation}
The sum on the right is a series of sidebands with amplitudes $J_m(a)$ and spacing $\delta_k$. Only one term in the sum can move at the resonant velocity, and for this term we explicitly write $m=r$, as in $J_r(a)$. Additionally, we will allow $a$ and $\delta_k$ to be functions of $z$ provided that they change slowly compared to the beat period ($2\pi/\delta_k$). In this case we will define $\psi_p=\int_0^z\delta_k(s)ds$ so that the instantaneous wavenumber can be broken into modes $k_g+m \delta_k(z)$.
%Note that we have ignored other modes of the DLA (corresponding to $k_g\rightarrow n k_g$) since these have phase velocities so far from $\beta_r$ that their contribution is negligible.  

Solving Maxwell's equations for each of these sidebands, we find that the Lorentz force is\,\cite{plettner_electromagnetic_2011}:
\begin{equation}
	\label{eq:force}
\begin{aligned}
    F_z &=i \sum_m q_e E_0 J_m(a) \cosh\left(\Gamma_m y \right) e^{i\left(\psi +(m-r)\left(\psi_p +\frac{\pi}{2}\right) \right)} \\
    F_y &=\sum_m q_e E_0 J_m(a) \frac{\Gamma_m}{k_m} \sinh\left(\Gamma_m y \right)  \left(1-\beta\beta_m\right) \times \\
    & \phantom{=-i \sum_m q_e E_0 J_m(a) \cosh\left(\Gamma_m y \right)} e^{i\left(\psi +(m-r)\left(\psi_p +\frac{\pi}{2}\right) \right)}
\end{aligned}
\end{equation}
where for simplicity we have assumed a perfectly symmetric mode and also that the diffraction efficiency is the same for each sideband. To complete this expression let us define a number of terms: firstly, we have the mode wavenumber $k_m=\partial_z (\psi + (m-r)\psi_p)$ with its associated phase velocity $\beta_m=\omega/(c k_m)$ and Lorentz factor $\gamma_m = \sqrt{1/(1-\beta_m^2)}$; then we have the evanescent factor $\Gamma_m = \frac{k_m}{\gamma_m}$; and finally we have the phase of a particle relative to the resonant mode $\psi=k_g z -\omega t +\psi_t $, where $\psi_t$ is a slowly changing phase we add to maintain synchronicity as the beam accelerates. It is also useful to define the normalized amplitude of each harmonic as: $\alpha_m= q_e E_0 J_m(a)/(m_e c^2 k_m)$. Note that $\alpha_m$ may be negative since $q_e$ and $J_m$ both have a sign; for example, in our design $\alpha_r>0$ and $\alpha_0<0$.

%Note that the force is not derivable from a potentional for the nonresonant harmonics and so niedermayer_beam_2017 is wrong!

%Naranjo et.al.[] create non-resonant harmonics by altering their dielectric structure so that it has a unit cell much longer than the $\beta \lambda$ needed to make the resonant mode. This creates sidebands like $k_\text{DLA} + n\delta_k$, and these sidebands are the nonresonant harmonic which provides pondermotive focusing. Rather than build the sidebands into our structure, we plan to program them in using an SLM. As illustrated in Fig.\,[1] we can apply a sinusoidal phase mask $\psi_p=a sin(k_\text{DLA} z /p)$ which splits the resonant DLA mode into a series of subharmonics, as seen via the Jacobi-Anger expansion:

%%%%%%%%%%%%%%%%%%%%%%%%%%%%%%%%%%%%%%%%%%%%%%%%%%%%%%%%%%%%%%%%%%%%%%%%%%%%%%%%%%
%%%%%%%%%%%%%%%%%%%%%%%%%%%%%%%%%%%%%%%%%%%%%%%%%%%%%%%%%%%%%%%%%%%%%%%%%%%%%%%%%%
%%%%%%%%%%%%%%%%%%%%%%%%%%%%%%%%%%%%%%%%%%%%%%%%%%%%%%%%%%%%%%%%%%%%%%%%%%%%%%%%%%
\subsection{Longitudinal dynamics}
	In our design the resonant harmonic dominates the time-averaged longitudinal dynamics. Thus, if we ignore the $y-z$ coupling in (\ref{eq:force}), the longitudinal dynamics is described (in a time-averaged sense) by the well-known pendulum Hamiltonian:
    \begin{equation}
\label{eq:long_hamiltonian}
	H_\parallel(\eta,\psi;k_r z) =\frac{\eta^2 }{2 \gamma_r^2 \beta_r^3} - \frac{\alpha_r}{\gamma_r}\left(\cos(\psi)+\psi\sin(\psi_r)\right)
\end{equation}
where $k_r z$ is the independent variable, $\psi$ is the particle phase, and  $\eta = (\gamma-\gamma_r)/\gamma_r$ is the fractional energy deviation. This Hamiltonian has a stable fixed point at the resonant phase $\psi_r$ for which the beam is linearly accelerating like $\partial_z\gamma_r = -\alpha_r k_r \sin(\psi_r)$. Surrounding the fixed point is a separatrix whose boundaries can be calculated as (c.f.\,\cite{kroll_free-electron_1981}):
\begin{equation}
	\label{eq:bucket_intercepts}
    \centering
  \begin{aligned}
      \cos(\psi_1)+\psi_1\sin(\psi_r)&=\cos(\psi_2)_+\psi_2\sin(\psi_r) \\
      \psi_2 &= \pi \, \text{sign}(\psi_r)-\psi_r
  \end{aligned}
\end{equation}
\begin{equation}
	\label{eq:bucket_height}
    \eta_\text{max}=\sqrt{2 \alpha_r \gamma_r \left(\cos(\psi_r)-\cos(\psi_2)+\left(\psi_r-\psi_2\right)\sin(\psi_r)\right)}
\end{equation}
Where $\psi_1$,$\psi_2$ are the locations where the bucket touches $\eta=0$, and $\eta_\text{max}$ is the height of the bucket at $\psi=\psi_r$. 

In addition to the time-averaged motion described by the pendulum Hamiltonian, the particles will undergo fast oscillatory motion due to the non-resonant forces. Taking the real part of (\ref{eq:force}) we can write the oscillatory force as $\partial_z \gamma =-\alpha_m k_m \sin(\psi +(m-r)\left(\delta_k z +\frac{\pi}{2}\right))$ (where we assume $\delta_k$, $a$, and $\psi_t$ are constants over the period of oscillation). To first order $\psi$ is also constant and the so the energy will oscillate. For $(m-r)=1$ the result would be:
\begin{equation}
	\label{eq:long_quiver}
    \gamma_\text{quiver} =- \alpha_m \frac{k_m}{\delta_k}\left(\sin(\psi+\left(\delta_k z\right))-\sin(\psi)\right)
\end{equation}

%%%%%%%%%%%%%%%%%%%%%%%%%%%%%%%%%%%%%%%%%%%%%%%%%%%%%%%%%%%%%%%%%%%%%%%%%%%%%%%%%%
%%%%%%%%%%%%%%%%%%%%%%%%%%%%%%%%%%%%%%%%%%%%%%%%%%%%%%%%%%%%%%%%%%%%%%%%%%%%%%%%%%
%%%%%%%%%%%%%%%%%%%%%%%%%%%%%%%%%%%%%%%%%%%%%%%%%%%%%%%%%%%%%%%%%%%%%%%%%%%%%%%%%%
\subsection{Buncher}
\label{sec:buncher}
%subscript $alpha$ to indicate mode?
%what is our buncher
	%compare to RFQ? %what does a buncher do? %focusing? %Cite narnjo and APF
	A buncher takes an unmodulated beam and redistributes its electrons so that they are grouped at a particular phase relative to the TM mode. The RFQ accomplishes this by using a weak resonant harmonic to execute roughly half a synchrotron oscillation before increasing the resonant phase to accelerate the bunched beam, but for a DLA the non-resonant harmonics are so strong that they disturb this gentle bunching. Instead, we can start our accelerator with the resonant harmonic turned off and use the oscillating field of the non-resonant terms to cause bunching\,\cite{naranjo_galaxie:_2013,niedermayer_alternating_2018}.
    
%Perturbation calculation 
	This bunching is a second order effect. We already saw in (\ref{eq:long_quiver}) that, to first order, the non-resonant harmonics just cause an oscillation of the beam energy. But some particles start at a phase which causes them to gain energy  while others start at a phase which causes them to lose energy, and those that gain energy will, on average, be moving a little faster than those which lose energy. This shows up as a second-order drift in the particle's phase, and so, ignoring fast oscillations and again using the case $(m-r)=1$, we find:
\begin{equation}
	\label{eq:bunching_phase}
	\psi^{(2)}=\psi+\frac{\alpha_m}{\gamma_r^3\beta_r^3} \frac{k_m}{\delta_k}\sin(\psi) z
\end{equation}
    The drift leads to bunching when $\partial_{\psi} \psi^{(2)}=0$. For negative $\alpha_m$ the bunching is centered around $\psi=0$, which is a perfect location to be trapped and then accelerated by the resonant harmonic. When bunched the energy spread will be exactly the quiver amplitude, $\alpha_m \frac{k_m}{\delta_k}$, and so we see that there is trade-off between the buncher length and the amount of heating imparted by the buncher.

%%%%%%%%%%%%%%%%%%%%%%%%%%%%%%%%%%%%%%%%%%%%%%%%%%%%%%%%%%%%%%%%%%%%%%%%%%%%%%%%%%
%%%%%%%%%%%%%%%%%%%%%%%%%%%%%%%%%%%%%%%%%%%%%%%%%%%%%%%%%%%%%%%%%%%%%%%%%%%%%%%%%%
%%%%%%%%%%%%%%%%%%%%%%%%%%%%%%%%%%%%%%%%%%%%%%%%%%%%%%%%%%%%%%%%%%%%%%%%%%%%%%%%%%
\subsection{Transverse Dynamics}
	The transverse force of (\ref{eq:force}) depends on $\sinh(\Gamma_m y)$ which we will linearize to $F_y\propto \Gamma_m y$. If we ignore the coupling to $z$ then it is natural to consider the oscillator strength $\kappa$ defined by $y'' = -\kappa_m y$ such that:    
\begin{equation}
	\label{eq:kappa}
    \kappa_m=-\frac{\alpha_m \cos(\psi +(m-r)\left(\psi_p(z) +\frac{\pi}{2}\right)) k_m^2}{\gamma_r \beta_r^2}\left(1-\beta_r\beta_m\right)
\end{equation}
Since the force is derivable from a potential we can use an analogy of Earnshaw's theorem to see that if the resonant harmonic is providing longitudinal focusing ($-\pi/2<\psi<\pi/2$) then it also has to be transversely defocusing. And indeed when $m=r$ the oscillator strength is $\kappa_r=-\frac{\alpha_r}{\gamma_r^3\beta_r^2} \cos(\psi) k_r^2$ which is defocusing at the resonant phase (i.e. $\psi=-\pi/3$).

	To overcome this resonant transverse defocusing we use the pondermotive motion caused by the non-resonant harmonics. This yields a term (the $(B+D)$ term of equation (3) in Ref.\,\cite{naranjo_stable_2012}) which is phase independent, with oscillator strength:
\begin{equation}
\label{eq:kappa_pondermotive}
\begin{aligned}[c]
	\kappa_p=\sum_{m\neq r}{\left(\frac{\alpha_{m}}{\gamma_r \beta_r^2}\right)^2  \left(\frac{k_m^4}{2 (m-r)^2 \delta_k^2}\right)}\,\, \times &\\
    \left(\left(1-\beta_r\beta_m\right)^2 +\frac{\left(1-\beta_r\beta_m\right)}{\gamma_r^2} \right)
    \end{aligned}
\end{equation}
The total oscillator strength is $\kappa=\kappa_p+\kappa_r$, which can be focusing for all phases if $|\kappa_p|>|\kappa_r|$. It is often useful to use the expansion $\left(1-\beta_r\beta_m\right)\approx\left(\frac{1}{\gamma_r^2}+\beta_r^2\frac{(m-r) \delta_k}{k}\right)$. Importantly, this expansion shows that at high energy $\kappa_p \propto \gamma_r^{-2}$ while $\kappa_r\propto \gamma_r^{-3}$ so that this scheme can be stable in the ultrarelativsitic limit.

Once the motion is transversely stable there remains a question of what the angular acceptance will be. The outermost oscillator will start (at rest) on the boundary and swing to have a maximum angle of $y'_\text{max}=\sqrt{-\kappa} y_\text{max}$. Only thus far, we haven't considered the fast oscillatory motion. Just like in the longitudinal case (\ref{eq:long_quiver}),  we can calculate the amplitude of small oscillations by assuming $\psi$ and $y$ are approximately constant. We find that under the influence of harmonics $m\neq r$ the rapid motion is described by:
\begin{equation}
	\label{eq:transverse_quiver}
    y_\text{quiver}=y\frac{\kappa_m}{(m-r)^2\delta_k^2}
\end{equation}
By construction this number should be small, but when calculating our aperture $y'_\text{max}$ we should account for the possibility of such motion by reducing $y_\text{max}$ commensurately.

%%%%%%%%%%%%%%%%%%%%%%%%%%%%%%%%%%%%%%%%%%%%%%%%%%%%%%%%%%%%%%%%%%%%%%%%%%%%%%%%%%
%%%%%%%%%%%%%%%%%%%%%%%%%%%%%%%%%%%%%%%%%%%%%%%%%%%%%%%%%%%%%%%%%%%%%%%%%%%%%%%%%%
%%%%%%%%%%%%%%%%%%%%%%%%%%%%%%%%%%%%%%%%%%%%%%%%%%%%%%%%%%%%%%%%%%%%%%%%%%%%%%%%%%
\section{Lattice Design}

%Describe the sections of the accelerator
Armed with our knowledge of the approximate beam dynamics we will discuss the design of a 2\,cm accelerator based on parameters available at the Pegasus facility.  Since we start with an unbunched beam our accelerator will need to consist of several sections to capture, bunch, and accelerate the beam. The prototype for such a linear accelerator, the RFQ, consists of 4 subsections: a radial matcher, a shaper, a buncher, and an accelerator. In order to limit our accelerator to a manageable 2\,cm we combine stages 1-3 into a rapid buncher. This causes us to have a reduced dynamic aperture and also heats the beam. Nonetheless, we consider this design suitable for a proof-of-principle experiment. 

\begin{figure}
\includegraphics[width=\columnwidth]{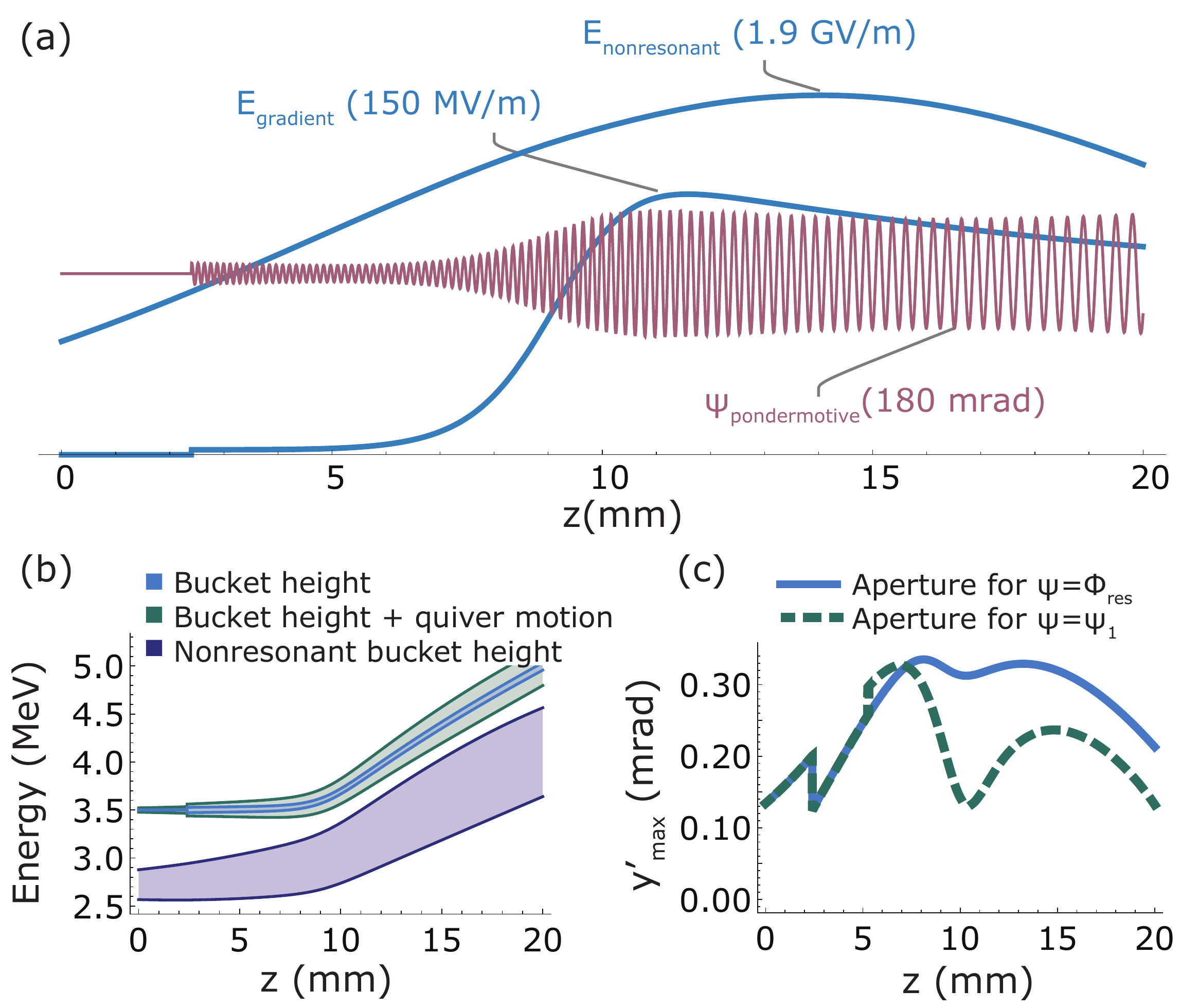}%
\caption{DLA lattice: (a) Curves showing the amplitude of the gradient, nonresonant harmonic, and the pondermotive portion of the phase mask. (b) An approximate longitudinal stability criterion: particles trapped in the resonant potential can't overlap with the nonresonant potential. (c) An approximate transverse stability criterion: we plot the maximum amplitude of Betatron oscillations (negative values would be unstable) }%
\label{fig:Lattice}
\end{figure} 

%Describe the free parameters
	Based on the phase mask discussed in section\,\ref{sec:theory} we have three free parameters available to us to both maintain stability and control the progression between stages of the accelerator. They are the mode spacing $\delta_k$, the mode amplitude $a$, and the phase $\psi_t$ which is used to control the resonant phase of the accelerator, $\psi_r$. The mode spacing can be changed without effecting the behavior of the resonant mode, and so we will use it to control the pondermotive focusing. The mode amplitude lets us control the ratio of the resonant to non-resonant harmonics, and so we will use it to regulate the effect of the resonant harmonic. Finally we will use resonant phase to control the progression of the accelerator from capture to acceleration. Once we have progressed through the accelerator and choosen values for these parameters we will be ready to integrate the equations of motion.

%Set E_0 and laser w
 Before choosing values for these free parameters we need to define the (unmodulated) gradient of the structure we are using. Based on a slightly optimistic reading of previous measurements\,\cite{cesar_high-field_2018} we will model an accelerator which can diffract 2\,GV/m into the first Bloch mode ($E_0=2$\,GV/m). But after losses in the optical transport of Fig.\,(\ref{fig:cartoon}), we will only have 8\,mJ of Ti:Sapphire (wavelength $\lambda$ = 800\,nm) with which to illuminate a structure of 2 cm length in $z$, and so we choose to focus the laser to a spot size of $w_x\times w_z=40\,\mu$m$\times 1.3$\,cm. This laser spot under-fills the structure (see Fig.\,(\ref{fig:Lattice}a)), but the leading edge of the Gaussian is sufficient to power the buncher and then the peak of the Gaussian can be used to accelerate.

%Explain the sidebands
Next we will choose which of the sidebands from (\ref{eq:Jacobi_Anger}) will be the resonant harmonic (the one we've been calling $r$). We will need to use most of the power for pondermotive focusing, so we should make $J_r(a)$ small. To accomplish this we choose $r=-1$ and use values of $a<<1$ so that most of the energy is in $J_0$. This choice is convenient because we can think of the motion as being due to just two terms: the $m=0$ and $m=-1$ harmonics; with the power splitting controlled by $a(z)$. We choose $r=-1$ instead of $r=1$ so that the non-resonant harmonic is slower than the resonant harmonic and $\kappa_p$ is focusing for all $\gamma$.

%Buncher + initial focusing
Now we are in a position to design the buncher. We set $a=0$ to put all the power into the non-resonant harmonic and start the second-order bunching. As discussed in section\,\ref{sec:buncher} the bunching will occur around $\psi=0$ (for $(m-r)=-1$) and will cause heating inversely proportional to the length of the bunching section. Since our buncher is also a `radial-matcher' we also need this section to provide strong focusing. From (\ref{eq:kappa_pondermotive}) we can can find that the pondermotive focusing favors small $\delta_k$ (for suitably low resonant energy). Thus we choose $\delta_k(0)\approx k_g/150$ which sets the initial transverse acceptance at about 0.15\,mrad (see Fig.\,(\ref{fig:Lattice}c)) and causes the beam to bunch in about 2.5\,mm. 

%Taper res phase and choose a
Immediately after bunching we increase $a$ (seen as a discontinuity in the plots of Fig.\,(\ref{fig:Lattice})) to establish a resonant harmonic at $\psi_r$=0, where the beam has bunched and where the seperatrix is largest. The initial value of $a$ is set to make the bucket height (\ref{eq:bucket_height}) more than large enough to include the entire energy spread. Then we slowly, but arbitrarily,  increase $\psi_r$ to turn on an accelerating gradient as shown in Fig.\,(\ref{fig:Lattice}a). Increasing $\psi_r$ would tend to shrink the bucket, so to keep the particles trapped we gradually increase $a$ in order to keep the bucket height constant. After $\approx1\,$mm we reach the maximum accelerating gradient of 150\,MV/m at $a$ = 0.18\,mrad and $\psi_r$ =$-\pi/3$\,rad. At this point we have defined enough free parameters to calculate the resonant energy:
\begin{equation}
\label{eq:energy_gain}
\gamma_r = \int_0^z{\frac{q_e E_0(s)}{m_e c^2}\sin(\psi_r(s))J_{-1}(a(s)) ds}
\end{equation}

%choose delta_k   & stability conditions
The only remaining task is to choose $\delta_k$ to stabilize the trajectories. We want to decrease $\delta_k$ in order to increase the pondermotive focusing, but doing so also makes the phase velocity of the non-resonant harmonic closer to the resonant velocity. At some point the `bucket' associated with the non-resonant harmonic will start to capture particles and the longitudinal motion will become unstable. According to the Chirikov criterion\cite{naranjo_stable_2012,chirikov_research_1971} the onset of this instability occurs when a particle undergoing a combination of synchrotron motion (\ref{eq:bucket_height}) and quiver motion (\ref{eq:long_quiver}) crosses into what would be part of the non-resonant potential's seperatrix. Fig.\,(\ref{fig:Lattice}b) shows how we toe this line in order to maximize the transverse stability shown in Fig.\,(\ref{fig:Lattice}c).

%calc taper phase
In order to complete the design we have to go back and calculate the taper phase, $\psi_t$, which is used to keep the resonant particle synchronous with the same sideband throughout the accelerator. To do so we track a test particle sitting at ($\psi=\psi_r(z)$, $y=0$). By construction it will have energy $\gamma_r$ calculated from (\ref{eq:energy_gain}) and knowing this we can calculate its time of arrival, $t_r$ at position $z$ as $c t_r=\int_0^z{(1/\beta_r)ds}$. Then we construct $\psi_t$ to keep the total phase in (\ref{eq:force}) constant: $\psi_t=\omega t_r -k_g z + \psi_r$.

\begin{figure}
\includegraphics[width=\columnwidth]{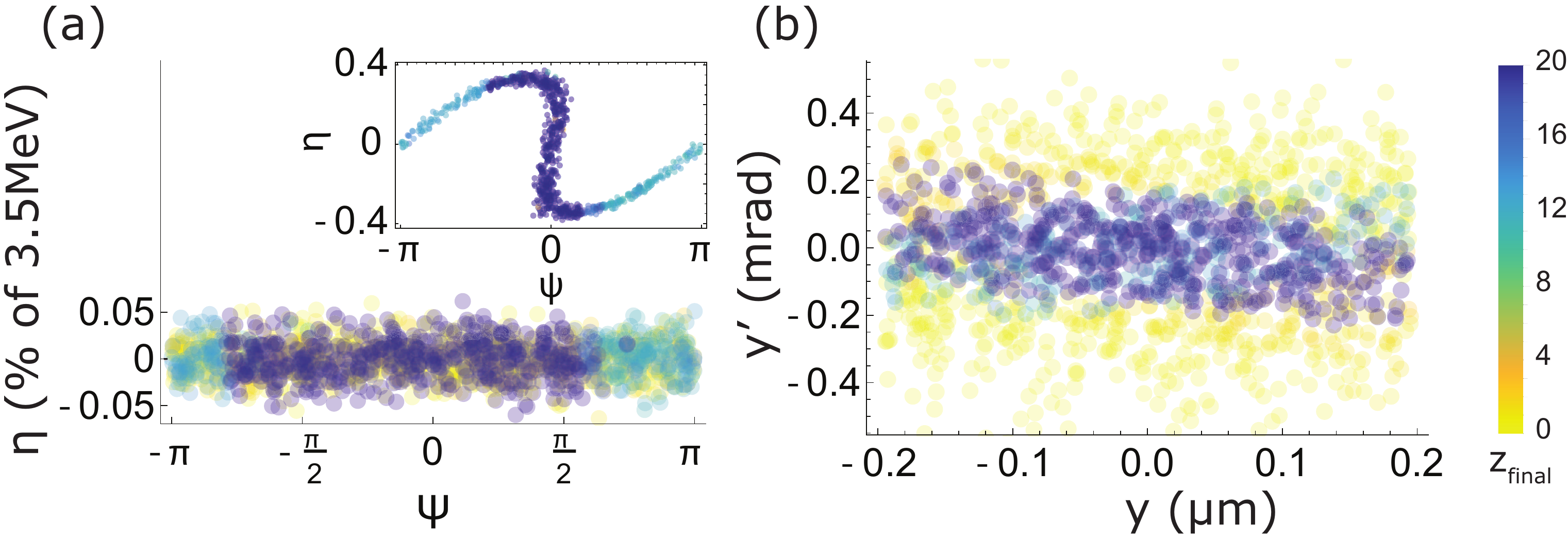}%
\caption{Dynamic aperture: (a) longitudinal phase space and (b) transverse phase space at the input to the DLA. Particles are color-coded by how far they make it into the accelerator (up to a maximum of 2\,cm). The inset to (a) shows a phase portrait after the buncher ($z$=2.5\,mm).}%
\label{fig:dynamic_aperture}
\end{figure}
%%%%%%%%%%%%%%%%%%%%%%%%%%%%%%%%%%%%%%%%%%%%%%%%%%%%%%%%%%%%%%%%%%%%%%%%%%%%%%%%%%
%%%%%%%%%%%%%%%%%%%%%%%%%%%%%%%%%%%%%%%%%%%%%%%%%%%%%%%%%%%%%%%%%%%%%%%%%%%%%%%%%%
%%%%%%%%%%%%%%%%%%%%%%%%%%%%%%%%%%%%%%%%%%%%%%%%%%%%%%%%%%%%%%%%%%%%%%%%%%%%%%%%%%
\section{Dynamic Aperture}

%Explain the calculations
To simulate the performance of this accelerator design we track 1500 particles through the DLA under the forces of (\ref{eq:force}), without consideration of collective effects. The ensemble we sample from is representative of the Pegasus beam: it has a mean energy of 3.5\,MeV, it starts unbunched with negligible energy spread, and it has a normalized transverse emittance of 20\,nm. The vacuum gap of the DLA is only 0.4\,$\mu$m, so for simplicity we stop tracking particles which cross $|y| \geq 0.2\,\mu$m.

%Discuss beam throughput
The resulting phase space acceptance is illustrated by color-coding the initial phase space based on how far the particles travel (Fig.\,(\ref{fig:dynamic_aperture})). Most of the particles are lost instantaneously (yellow) because their initial angle carries them into the wall of the DLA. Of the rest there are two groups: those that get bunched (purple), and those that remain out of phase (blue), as shown in the inset. The bunched particles will end up seeing a net defocusing force, while the opposite is true for the out of phase particles, thus the transverse acceptance for these two groups is slightly skewed (as can be seen in Fig.\,(\ref{fig:dynamic_aperture}b)). The out of phase particles are eventually lost when the resonant phase ramps up and the defocusing force becomes stronger.

In total, we capture 70\% of the particles that survive the first 1\,mm, and about 30\% of the all the particles shown in Fig.\,(\ref{fig:dynamic_aperture}). If we account for particles which never made it into the structure, either because they started outside the vacuum gap or because they were temporally mismatched (the electron beam is 100\,fs long, but the laser is only 60\,fs long) then we get a total transmission on the order of 0.1\%. Assuming an initial charge of 50\,fC we can expect roughly 20 electrons/bunch for 15 bunches. This number can be improved by several orders of magnitude if we use a flat beam transform to improve the 1D brightness\,\cite{ody_flat_2016} and if we velocity bunch the electron beam enough to eliminate the temporal mismatch\,\cite{maxson_direct_2017}.
%%%%%%%%%%%%%%%%%%%%%%%%%%%%%%%%%%%%%%%%%%%%%%%%%%%%%%%%%%%%%%%%%%%%%%%%%%%%%%%%%%
%%%%%%%%%%%%%%%%%%%%%%%%%%%%%%%%%%%%%%%%%%%%%%%%%%%%%%%%%%%%%%%%%%%%%%%%%%%%%%%%%%
%%%%%%%%%%%%%%%%%%%%%%%%%%%%%%%%%%%%%%%%%%%%%%%%%%%%%%%%%%%%%%%%%%%%%%%%%%%%%%%%%%

\section{Output phase space}
\begin{figure}
\includegraphics[width=\columnwidth]{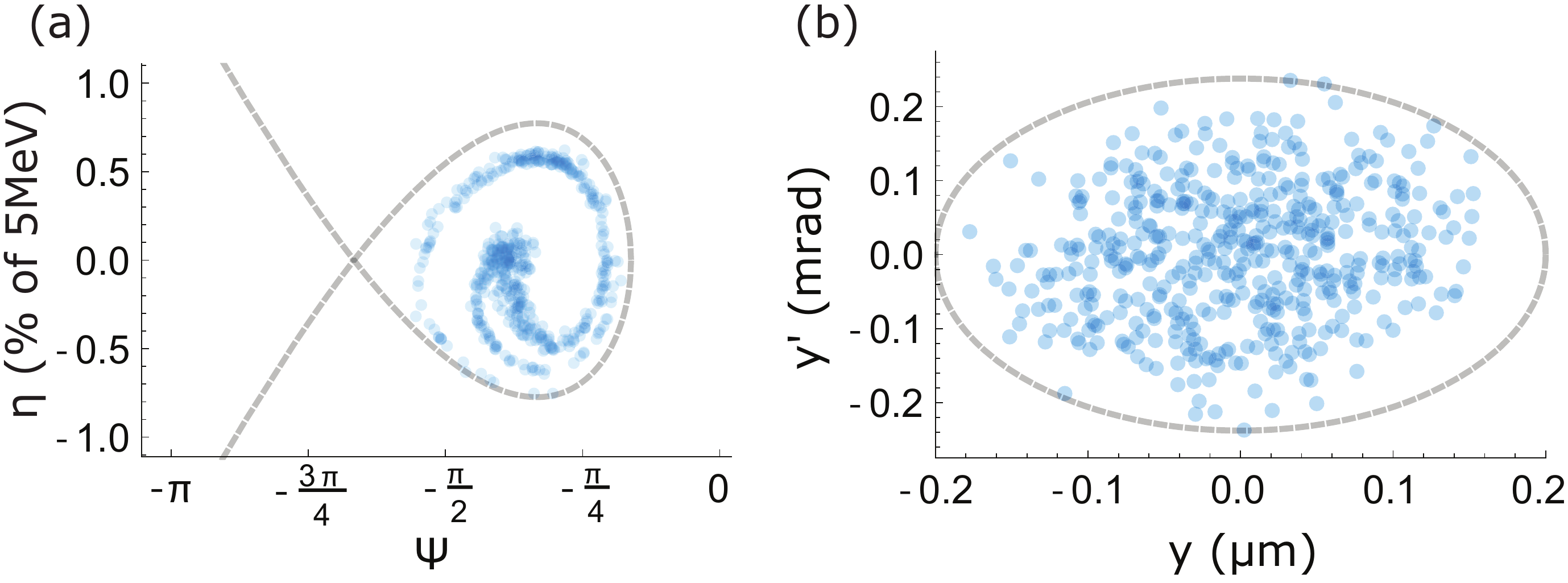}%
%\caption{Output phase space averaged over the last 0.4\,mm of the DLA: (a) longitudinal and (b) transverse. Dashed lines show the linearized buckets of \S\,\ref{sec:theory}}%
\caption{(a) longitudinal and (b) transverse phase-spaces averaged over the last 0.4\,mm of the DLA. Dashed lines show the linearized buckets of section\,\ref{sec:theory}}%
\label{fig:average_ps}
\end{figure}

We can compare the numerical solutions to the secular dynamics discussed in section\,\ref{sec:theory} by time-averaging the particle motion. In Fig.\,(\ref{fig:average_ps}) we show a phase space portrait of the beam averaged over the last 0.4\,mm of the DLA (the length of the final beat period, $2\pi/\delta k$). We can see that the longitudinal dynamics are very well described by the time averaged bucket, while the transverse dynamics appear more tightly bound (in $y$) than might be expected since the quiver motion kills particles near the boundary. 

If we were to show an instantaneous phase portrait then the non-resonant harmonics would significantly alter the distribution. For example, the transverse dynamics has a large $y-y'$ correlation which oscillates as the nonlinear harmonics slip by. A similar effect happens longitudinally, where the beam centroid oscillates up and down relative to $\eta=0$ and also the distribution gets sheared by nonlinear focusing. Nonetheless, the motion is stable and can be profitably regarded in the time-averaged sense as analogous to a conventional accelerator undergoing both synchrotron and betatron oscillations.

\section{Conclusions}
We have outlined the steps needed to design a DLA which first bunches and then accelerates a beam by 1.5\,MeV at a peak gradient of 150\,MeV/m. We use approximate models of the beam dynamics to describe the evolution of the bunch parameters and ensure the stability of the accelerating buckets. We show that a programmable phase array provides sufficient flexibility to turn a single DLA structure into a buncher, focusing channel, and accelerator. And although future DLAs may be based on alternative approaches\,\cite{wootton_dielectric_nodate,hughes_-chip_2018}, the tunability of this method is very valuable for controlling and studying beam dynamics in a DLA, and may inform future DLA designs.

%I dunno about this one....
%Due to the nanoscopic size of the DLA period we are able to fit a significant amount of beam dynamics into a 2\,cm accelerator. In terms of the time-averaged particle motion the beam undergoes about 4.6 synchrotron oscillations and 4.4 betatron oscillations. In this, or in slightly longer DLA's, we have the possibility of study tune resonances in a strongly non-linear, coupled accelerator. Additionally,  we have only considered a perfectly symmetric TM accelerating mode, but it is easy to make a DLA with a time-varying dipole moment which could be used to drive instabilities in the focusing channel. This makes DLA a compact platform for.... 


\bibliographystyle{IEEEtran}
\bibliography{IEEEabrv,./bib}


%\bibliographystyle{IEEEtran}

%\begin{thebibliography}{1}


%\end{thebibliography}

\balance

\end{document}